# Chromatic Adaptation Transform by Spectral Reconstruction

Scott A. Burns, University of Illinois at Urbana-Champaign, scottallenburns@gmail.com

February 28, 2019

**Note to readers**: This version of the paper is a preprint of a paper to appear in Color Research and Application in October 2019 (Citation: Burns SA. Chromatic adaptation transform by spectral reconstruction. *Color Res Appl.* 2019;44(5):682-693). The full text of the final version is available courtesy of Wiley Content Sharing initiative at: https://rdcu.be/bEZbD. The final published version differs substantially from the preprint shown here, as follows. The claims of negative tristimulus values being "failures" of a CAT are removed, since in some circumstances such as with "supersaturated" colors, it may be reasonable for a CAT to produce such results. The revised version simply states that in certain applications, tristimulus values outside the spectral locus or having negative values are undesirable. In these cases, the proposed method will guarantee that the destination colors will always be within the spectral locus.

*Abstract: A color appearance model (CAM) is an advanced colorimetric tool used to predict color appearance under a wide variety of viewing conditions. A chromatic adaptation transform (CAT) is an integral part of a CAM. Its role is to predict "corresponding colors," that is, a pair of colors that have the same color appearance when viewed under different illuminants, after partial or full adaptation to each illuminant. Modern CATs perform well when applied to a limited range of illuminant pairs and a limited range of source (test) colors. However, they can fail if operated outside these ranges. For imaging applications, it is important to have a CAT that can operate on any real color and illuminant pair without failure. This paper proposes a new CAT that does not operate on the standard von Kries model of adaptation. Instead it relies on spectral reconstruction and how these reconstructions behave with respect to different illuminants. It is demonstrated that the proposed CAT is immune to some of the limitations of existing CATs (such as producing colors with negative tristimulus values). The proposed CAT does not use established empirical corresponding-color datasets to optimize performance, as most modern CATs do, yet it performs as well as or better than the most recent CATs when tested against the corresponding-color datasets. This increase in robustness comes at the expense of additional complexity and computational effort. If robustness is of prime importance, then the proposed method may be justifiable.*

# 1. Introduction

A color appearance model (CAM) is an advanced colorimetric tool used to predict color appearance under a wide variety of viewing conditions.[1,2] A chromatic adaptation transform (CAT) is an integral part of a CAM. Its role is to predict "corresponding colors," that is, a pair of colors that have the same color appearance when one is viewed under one illuminant and the other is viewed under a different illuminant.[3] It is assumed that the viewer is adapted, or partially adapted, to the illuminants before assessing color appearance.

Most CATs are based on the von Kries coefficients law.[4] The tristimulus values of a source sample are transformed to cone-response space (or something similar, such as a "sharpened" version), and then the cone response magnitudes are scaled by the ratios of destination-to-source illuminant white point cone-response magnitudes. A final transformation back to tristimulus value space yields the predicted destination color. Additional parameters are usually added to account for partial adaptation.

Note that the terms "source" and "destination" used here differ from the alternate terminology of "test" and "reference" often found in the literature. If the latter terms are preferred, simply replace "source" with "test" and "destination" with "reference" in the development that follows.

Several experimental datasets of corresponding colors have been compiled to aid in the development of modern CATs. The standard datasets used in the development of recent CATs are known as CSAJ, Helson, Lam & Rigg, LUTCHI, Kuo & Luo, Breneman, Braun & Fairchild, and McCann.[5] The numerical data are no longer available at the URL provided in Reference 5, but they can be obtained through a web archive site.[6]

Some von-Kries-type CATs are theoretical, that is, they do not rely on the experimental corresponding-color datasets for their development. Instead, they are derived from the results of well-established physiological studies, such as spectral cone-sensitivity functions or cone signal compression phenomena. In contrast, many recent CATs are derived with the aid of the empirical corresponding-color datasets. The transformation that converts tristimulus values to cone responses is modified through an optimization to yield a CAT that is able to match the empirical data more closely. The optimization sometimes includes constraints to avoid some performance issues, for example, ensuring that the destination tristimulus values are always positive for "popular" source/destination illuminant combinations and that it obeys the "nesting" rule.[7-13] Examples of these empirically "trained" CATs include Bradford (BFD), CMCCAT97, CMCCAT2000, CAT02, and most recently CAT16.[13-15]

In the sections that follow, several examples are given of modern CATs performing poorly, specifically, yielding negative destination tristimulus values, or destination colors outside the spectral locus, for reasonable combinations of source/destination illuminants and source colors. Then a more robust chromatic adaptation transformation is proposed that makes use of a spectral reconstruction technique. The proposed method does not use experimental corresponding-color datasets in its development, yet performs as well as or better than all existing modern CATs when tested against the experimental datasets. It is not susceptible to some problems existing CATs experience, even for the most extreme of source/destination illuminants and source colors. The penalty for this improved robustness, however, is that it requires considerably more computational effort, which may limit its suitability for some applications.

## 2a. Limitations of Existing CATs—Nonstandard Source Illuminants

Recent CATs are effective in predicting corresponding colors within certain limits. First, they usually assume that the destination illuminant is similar to daylight (such as equal-energy, D65, or C). Second, the source illuminant is usually assumed to be similar to commonly used light sources (tungsten, fluorescent, white LED, etc) and not extreme sources like gas discharge lamps (neon, metal halide, sodium vapor, etc.).

To demonstrate the performance of CAT02 with appropriate illuminants, reflectance curves for 1485 Munsell color chips (2007 Glossy) were converted to tristimulus values using illuminant A, and then plotted on the chromaticity diagram. They are shown in the left part of Figure 1. These tristimulus values were then processed by CAT02 (formulated to convert from illuminant A to C with full adaptation, $D = 1$). The results are plotted on the chromaticity diagram in the right part of Figure 1. The distribution of transformed colors seems very reasonable; no obvious defects stand out. The distributions remain within the spectral locus and are clustered around the illuminant white points, as expected.

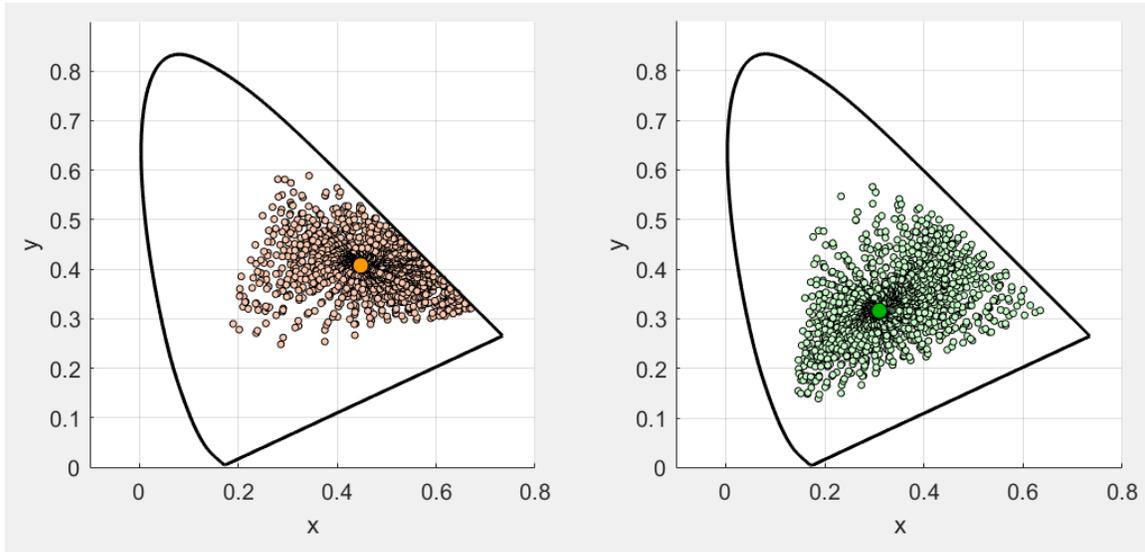

Figure 1. Chromaticity coordinates of Munsell reflectance data (2007 Glossy) computed with illuminant A (left) and after the transformation by CAT16 from illuminant A to C (right). Large dots are illuminant white points.

In contrast, consider what happens when the source illuminant is changed from illuminant A to the spectral distribution shown in Figure 2. This purplish-blue illuminant has white point $(X, Y, Z) = (2.15, 1, 4.73)$ and white point chromaticity $(x, y) = (0.272, 0.127)$. Figure 3 shows the source locations of the Munsell colors with this new source illuminant (left) and the destination locations as transformed by CAT02, using illuminant C as the destination illuminant (right). The large dots in each figure are the illuminant white points.

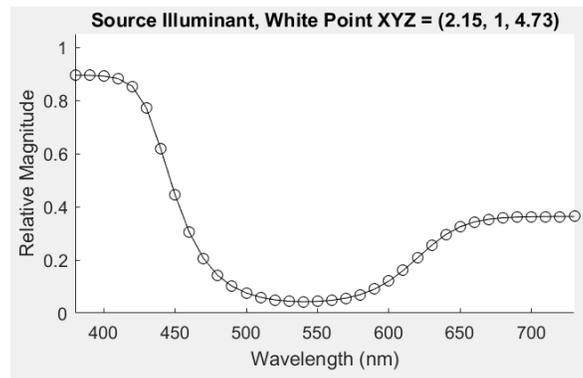

Figure 2. Source illuminant for results shown in Figure 3.

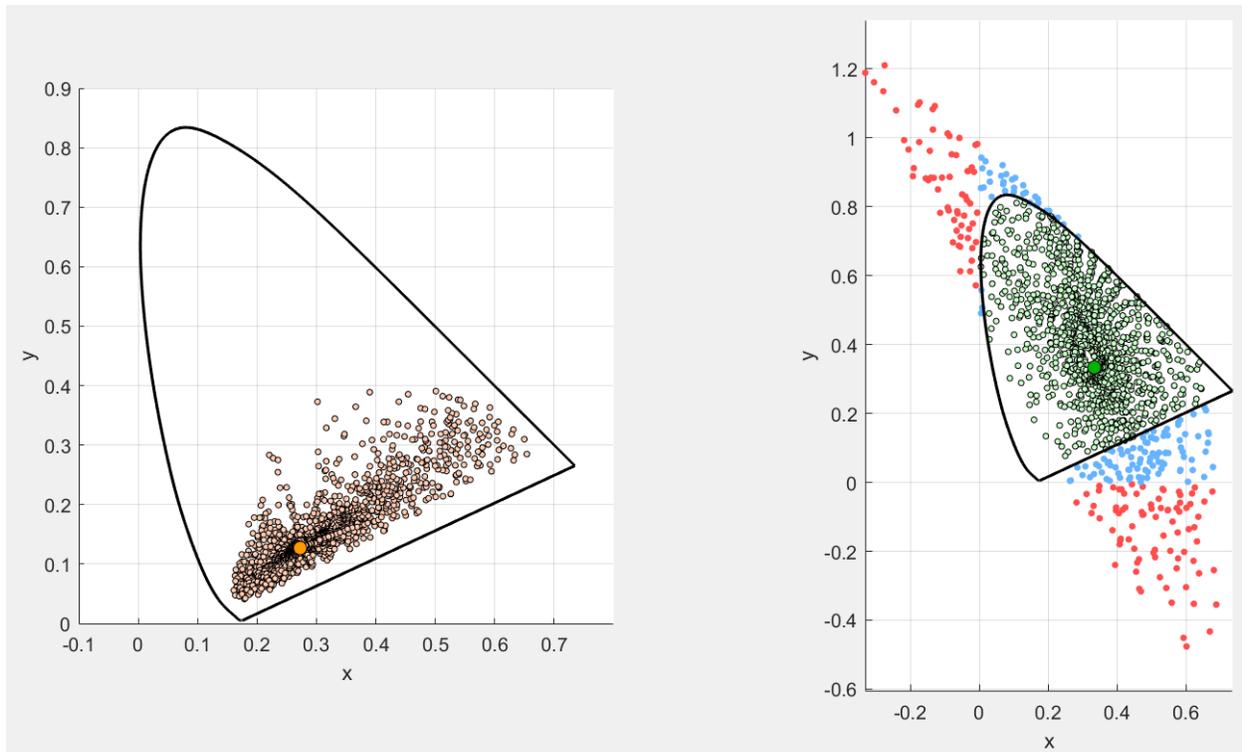

Figure 3. Chromaticity coordinates of Munsell reflectance data (2007 Glossy) computed with the illuminant in Figure 2 (left) and after the transformation by CAT02 to illuminant C (right). Red dots have negative tristimulus values and red and blue dots are outside the spectral locus. Large dots are illuminant white points.

It is evident that CAT02 performs poorly; many of the Munsell colors are transformed to points outside the chromaticity diagram (blue and red). The red ones also have negative destination tristimulus values. A numerical example of one instance of negative tristimulus destination values is presented in the online supplementary documentation.[16] The shortcomings of CAT02 has been well documented in the literature.[7-11] The poor performance (when using Munsell source colors) is not shared by CAT16, which was optimized specifically to avoid negative tristimulus values. The same is true for the "untrained" CAT known as HPE.[17-18] This CAT is a direct application of the von Kries coefficient rule using a transformation to cone response space derived from Estevez's original estimates of theoretical cone sensitivity functions.[19] HPE does not use the experimental corresponding-color datasets for improved performance and transforms the Munsell colors to destinations within the spectral locus for even the most extreme source illuminants, with the caveat that the destination illuminant is not too far from equal energy. A later section of this paper will describe the limitations of HPE, CAT02, and CAT16 when the destination illuminant moves away from equal energy. But before doing that, it is instructive to note that all three of these CATs can produce negative destination tristimulus values when applied to certain "object colors" even when the destination illuminant is equal energy, as demonstrated in the next section.

**2b. Limitations of Existing CATs—Application to All Object Colors as Source Colors**

If we restrict our attention to objects having reflectance properties in the range 0 to 1 over all visible wavelengths, then the stimuli produced by a given illuminant acting on all possible reflectances of this type project into a three-dimensional color space region called the "object color solid," the outer surface of which comprises the "optimum colors." [20] Testing a CAT with all object colors is more demanding than using just the Munsell colors, since the latter falls well within the interior of the object color solid. It will be demonstrated that HPE, CAT02, and CAT16 all produce negative destination tristimulus values for some portions of the object color solid, when used as source colors. This happens even for source illuminants that are not very extreme (i.e., not having chromaticities that approach the spectral locus), while using equal energy as the destination illuminant. Figure 4 shows HPE, CAT02, and CAT16 applied to a $Y = 0.3$ slice of the object color solid, using a source illuminant that is halfway between equal energy and the spectral locus (white point $XYZ = (1.5, 1, 0.5)$ and $xy = (0.5, 0.333)$, shown in Figure 5).

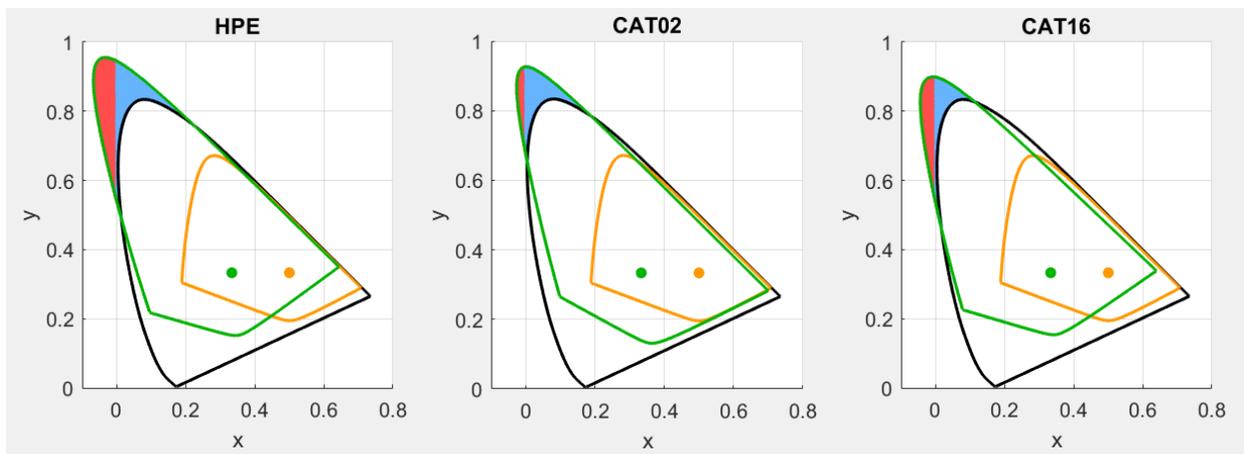

Figure 4. Object color solid slice for $Y = 0.3$ (orange curve) using source illuminant shown in Figure 5 (orange dot) and the transformed region (green curve) after application of HPE, CAT02, and CAT16, using equal energy as the destination illuminant (green dot). The red region has negative tristimulus values and the red and blue regions fall outside the spectral locus.

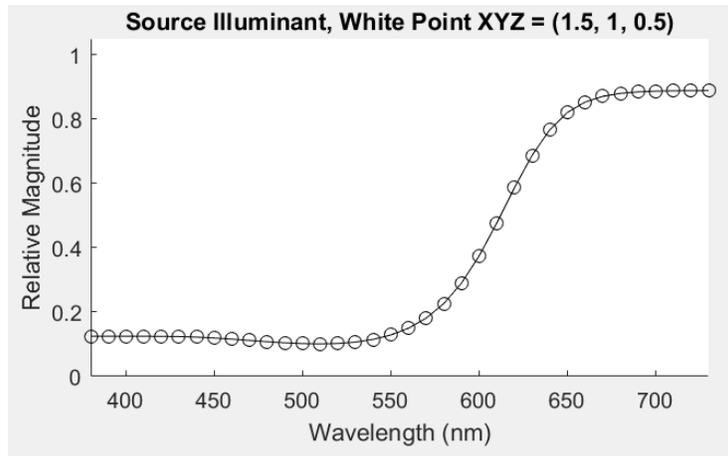

Figure 5. Source illuminant for the results shown in Figure 4.

A substantial portion of the object colors are transformed outside of the spectral locus (red and blue regions) and the red region has negative destination tristimulus values. These examples demonstrate that modern CATs (and even a standard, untrained von Kries CAT like HPE) can give negative destination tristimulus values for some object colors. However, as long as the destination illuminant is similar daylight or equal energy, the cases of negative destination values seem to happen mostly with source tristimulus values having relatively low $Y$ values, such as $Y = 0.3$ in example above.

**2c. Limitations of Existing CATs—Nonstandard Destination Illuminants**

Some CATs have specifications that assume the destination illuminant is equal energy and most of the correspond-color datasets have destination illuminants of D65 or D55, which are in the vicinity of equal energy. This section examines what happens when a more chromatic destination illuminant is used. Figure 6 demonstrates that if equal energy is used as the *source* illuminant, and a greenish illuminant is used as the destination (white point $XYZ = (0.660, 1, 0.792)$ and $(x, y) = (0.269, 0.408)$, shown in Figure 7), then many of the Munsell colors are transformed out of the spectral locus by HPE, CAT02, and CAT16, and a good portion of those have negative tristimulus values.

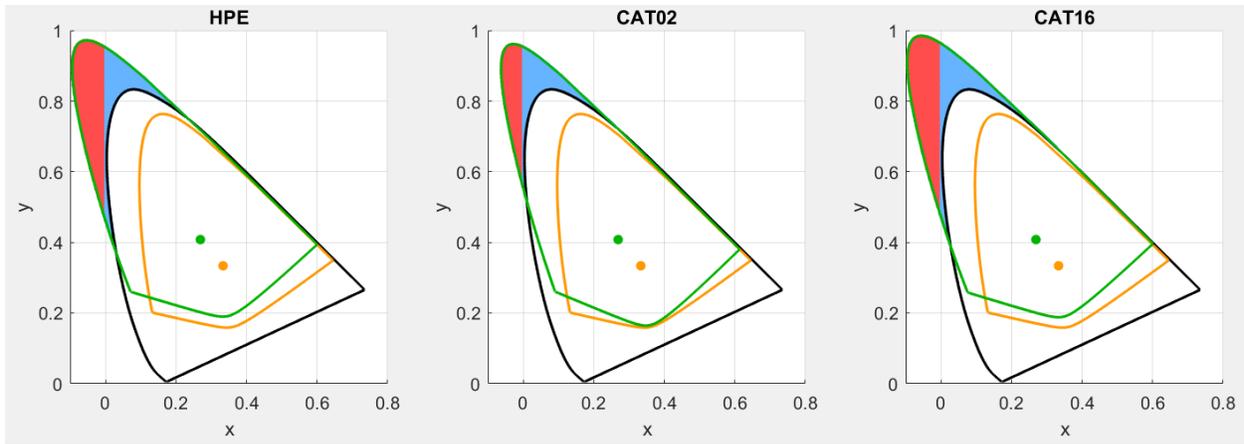

Figure 6. Object color solid slice for $Y = 0.3$ (orange curve) using equal energy as the source illuminant (orange dot) and the transformed region (green curve) after application of HPE, CAT02, and CAT16, using the illuminant in Figure 7 as the destination illuminant (green dot). The red region has negative tristimulus values and the red and blue regions fall outside the spectral locus.

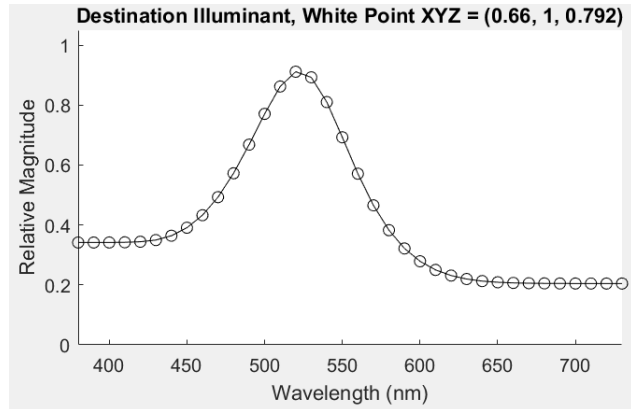

Figure 7. Destination illuminant for the results shown in Figure 6.

The shift of the destination illuminant from equal energy is only 20% of the distance to the spectral locus, yet significant portions of the object color solid transform to outside of the spectral locus.

In summary, modern CATs can be somewhat "fragile" and give poor results when the source/destination illuminants move too far away from commonly-used illuminants, or when source colors are selected too near the object color solid boundary. The following sections will present an alternative CAT that is based on spectral reconstruction. It does not use the empirical corresponding-color datasets to fine-tune its performance, as is the case with CAT02 and CAT16. It will be shown to perform as well as or better than CAT02 and CAT16 when applied to the corresponding-color datasets and to be immune to transforming source colors to destinations outside the spectral locus or to negative tristimulus values.

## 3. Notation

The symbols used in the following development are summarized in Table 1.

Table 1. Explanation of Symbols

| Symbol | Definition |
|---|---|
| $n$ | Number of wavelength bands used to discretize spectral-based entities. (In this presentation, all computations use $n = 36$, representing 10 nm bands spanning the range 380 nm to 730 nm.) |
| $\bar{x}, \bar{y}, \bar{z}$ | Three $n \times 1$ vectors of color-matching functions. (The CIE 1931 standard observer is used in computations in this presentation.) |
| $A$ | $n \times 3$ matrix of color-matching functions (CMF), by columns: $A = [\bar{x}, \bar{y}, \bar{z}]$. |
| $W$ | $n \times 1$ illuminant vector, assumed to be non-negative and scaled (normalized) so that the scalar product $\bar{y}' \cdot W = 1$. (Prime denotes transpose. Subscripts denote different illuminants, e.g., $W_S$ and $W_D$.) |
| $\overline{W}$ | $n \times n$ diagonal matrix, $\overline{W} = \mathrm{diag}(W$, which places $W$ on the main diagonal and zeros elsewhere. |
| $\rho$ | $n \times 1$ reflectance vector, typically with values in the range 0-1. |
| $XYZ$ | $3 \times 1$ vector of tristimulus values, $X, Y,$ and $Z$, computed as $XYZ = A' \overline{W} \rho$. (Note the reflective/transmissive case is used here, as opposed to the emissive case, so all tristimulus values are referenced to some illuminant, $W$, and a perfect reflector will yield $Y=1$. A subscript indicates to which illuminant it is referenced, e.g., $XYZ_S$ or $XYZ_D$. A second subscript of "$p$" indicates it is a predicted quantity generated by a CAT, e.g., $XYZ_{Dp}$.) |
| $XYZ^{wp}$ | $3 \times 1$ vector of illuminant white point tristimulus values, $XYZ^{wp} = A' W$. (Because of the assumed scaling of $W$, the $Y$ value will always equal 1. A subscript can be added to specify the illuminant, e.g., $XYZ_S^{wp}$ or $XYZ_D^{wp}$.) |
| $A_W$ | $n \times 3$ matrix of illuminant $W$-referenced color-matching functions, $A_W = \overline{W} A$. ($W$-referenced tristimulus values are computed directly from reflectance as $XYZ_W = A_W' \rho$. The subscript is used to differentiate between various illuminants, e.g., $A_S$ and $A_D$.) |
| $C$ | $n \times n$ matrix of finite-differencing constants (see Equation 5). |
| $A_{SD}$ | $n \times 3$ matrix of dual-referenced (by both source and destination illuminants) color-matching functions, $A_{SD} = \overline{W_S} \, \overline{W_D} \, A$. |
| $D$ | Degree of adaptation parameter, ranging from 0 (fully unadapted) to 1 (fully adapted). |

## 4. Chromatic Adaptation Transform by Spectral Reconstruction

Modern CATs ignore the spectral makeup of illuminants and source/destination colors and operate only on tristimulus values. Changes in color sensations by adaptation, ostensibly due to changes in cone sensitivities, are modeled by the von Kries coefficients law. In contrast, the proposed CAT constructs spectral distributions from the tristimulus values and simply computes the effect that changing the illuminant has on tristimulus values using standard CMF calculations. While this approach appears to ignore the phenomenon of adaptation, focusing instead on techniques more closely related to color constancy, the corresponding-color pairs it generates match the empirical corresponding-color datasets as well as or better than existing CATs.

The proposed CAT algorithm operates as follows: Given a source color, $XYZ_S$, which is referenced to a source illuminant with white point $XYZ_S^{wp}$, we seek a prediction of a corresponding color, $XYZ_{Dp}$, referenced to a destination illuminant with white point $XYZ_D^{wp}$. (At this point, we are assuming full adaptation.)

1. Construct two normalized spectral distributions, $W_S$ and $W_D$, which when treated as illuminants, have white points $XYZ_S^{wp}$ and $XYZ_D^{wp}$, respectively. (See a later section on how to perform this spectral reconstruction.) Use these two distributions to form $W_S$- and $W_D$-referenced CMFs, $A_S = \overline{W_S}\,A$ and $A_D = \overline{W_D}\,A$.
2. Construct a reflectance curve, $\rho$, that satisfies $XYZ_S = A_S{'}\,\rho$. (See a later section on how to perform this reflectance reconstruction.)
3. Compute $XYZ_D = A_D{'}\,\rho$.
4. Adjust $XYZ_D$ so that it preserves its chromaticity coordinates, $(x, y$, while having a $Y$ value that matches the $Y$ value of $XYZ_S$. Call this $XYZ_{Dp}$.
5. $XYZ_{Dp}$ is the predicted corresponding color associated with $XYZ_S$ for the two illuminant white points $XYZ_S^{wp}$ and $XYZ_D^{wp}$.

The fourth step above is justified by the observation that in corresponding color experiments, the goal of the observer is to find source and destination colors that appear identical when viewed under a state of adaptation to the respective illuminants. The experiments are designed to produce corresponding colors with similar luminances, so that luminance-dependent changes in colorfulness and contrast are not triggered (known as Hunt and Stevens effects). Indeed, 14 of the 26 standard corresponding-color datasets (the three LUTCHI sets, two Kuo & Luo sets, and nine Breneman sets) have corresponding color pairs with identical $Y$ values. The mean difference of all $Y$ values in the corresponding-color data (excluding McCann) is 0.33 (on a 0-100 scale) and the standard deviation of the differences is 1.1. Steps 1-3 of the algorithm above do not preserve $Y$ values as the corresponding-color experiments do, so it is reasonable to add step 4 to mimic this preservation.

Before examining the specifics of how to compute the spectral reconstructions above, a demonstration of the effectiveness of the proposed CAT is presented in Table 2, which compares the $\Delta E^*_{94}$ color differences between the experimental dataset destination colors, $XYZ_D$, and the CAT-predicted destination colors, $XYZ_{Dp}$, for four methods: HPE, CAT02, CAT16, and the proposed CAT. In all cases, full adaptation ($D = 1$) is used. See the online supplementary documentation for more information regarding these calculations.[21]

Table 2. Mean $\Delta E^*_{94}$ color difference for HPE, CAT02, CAT16, and the proposed method.

| Dataset | HPE | CAT02 | CAT16 | Proposed Method |
|---|---|---|---|---|
| CSAJ | 4.71 | 3.66 | 3.95 | 3.72 |
| Helson | 4.52 | 3.45 | 4.00 | 4.10 |
| Lam & Rigg | 4.31 | 2.97 | 3.45 | 3.22 |
| LUTCHI | 4.03 | 3.55 | 3.43 | 4.01 |
| Kuo & Luo | 4.29 | 3.30 | 3.41 | 2.85 |
| Breneman | 6.61 | 5.70 | 5.66 | 5.48 |
| Braun & Fairchild | 4.54 | 4.00 | 4.24 | 4.07 |
| McCann | 10.82 | 11.52 | 10.80 | 9.78 |
| **Mean (All Datasets)** | 5.54 | 4.87 | 4.91 | 4.74 |
| **Mean (No McCann)** | 4.77 | 3.91 | 4.06 | 4.01 |

The mean values in the last two rows of Table 2 are weighted means, that is, they represent the mean of all samples in all datasets, not the mean of the dataset means. The proposed method performs comparatively the best when all datasets are considered. Both CAT02 and CAT16 were optimized for all datasets except the McCann dataset. The proposed method falls between the performance of CAT02 and CAT16 when the McCann data is excluded. It should be noted that the proposed method, like the HPE method, is not optimized at all to fit the experimental corresponding-colors datasets. Such optimization could be applied to the proposed method to further improve its performance, but possibly at the expense of making it less robust, like CAT02 and CAT16.

One possible explanation for why the spectral reconstruction method performs so well is the observation that the type of reconstructed reflectance curve used in this study seems to match reflectance curves of object colors found in nature surprisingly well, in particular, the reflectances of common paints and pigments.[22] Figure 8 shows the spectrophotometrically measured reflectance curves of six Liquitex® paints and the corresponding reflectance reconstructions generated from the tristimulus values of the paints.

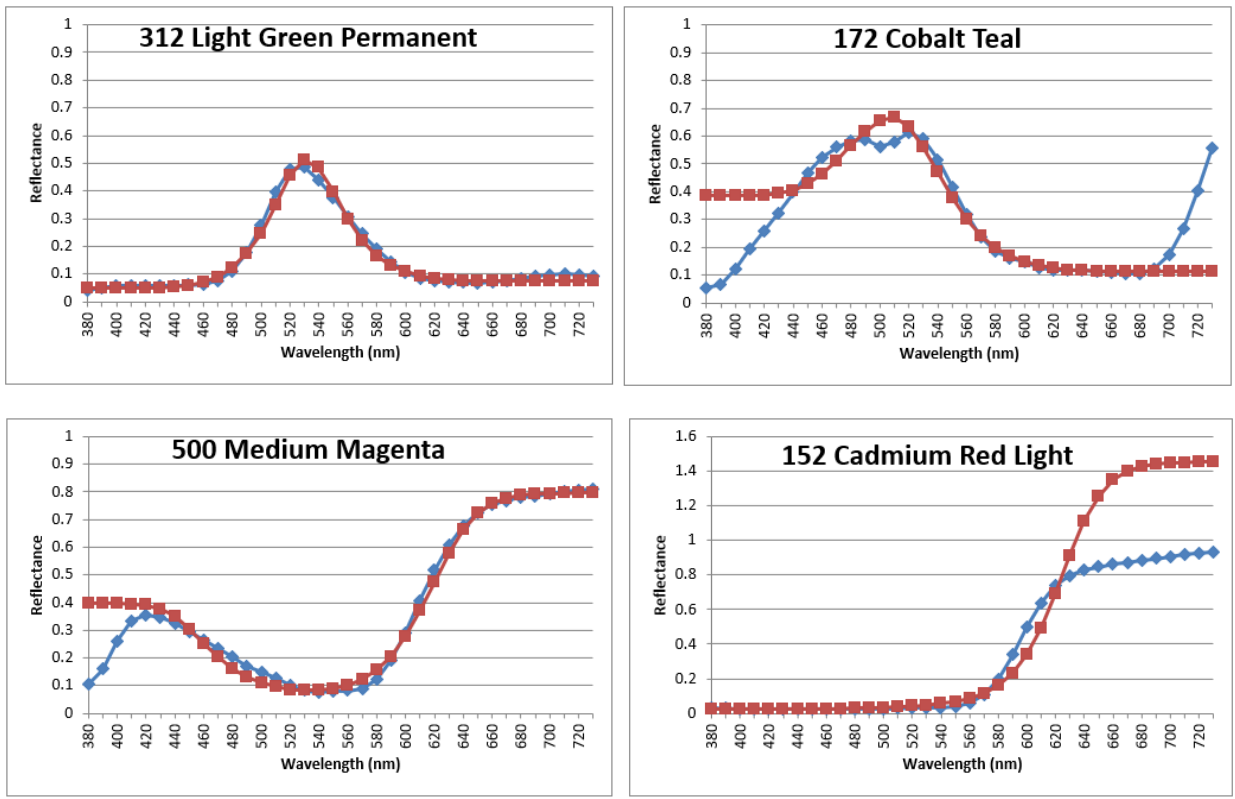

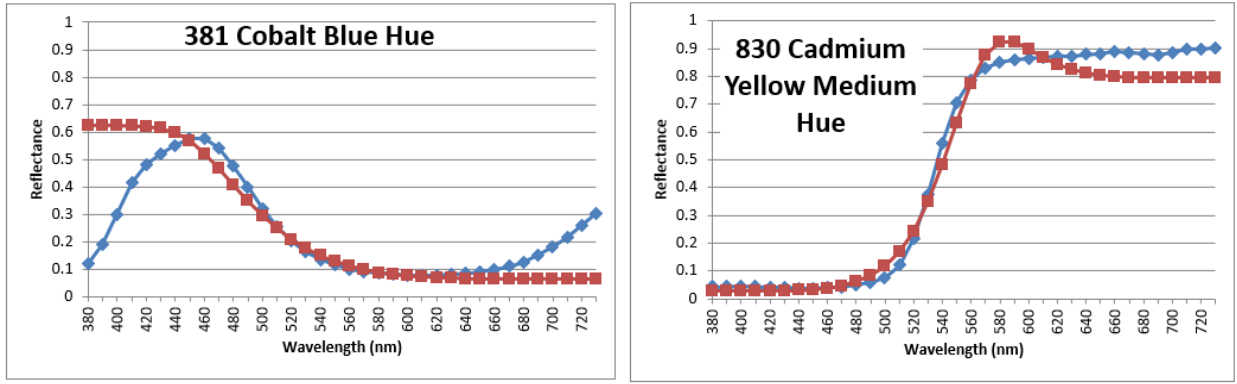

Figure 8. Spectrophotometrically measured reflectance curves of six Liquitex® paints (blue curves) and the corresponding reflectance reconstructions generated from the tristimulus values of the paints (red curves).

Although there are considerable discrepancies in the upper and lower extremes of the visible spectrum, these deviations are largely inconsequential, as human sensitivity to these wavelengths is very low. Consequently, calculation of quantities like tristimulus values are relatively insensitive to variations in stimuli at these extreme wavelengths.

## 5. Spectral Reconstruction Computation

Spectral reconstruction is the process of generating a distribution (e.g., reflectance, power, etc.) over a wavelength (or frequency) domain, given only a three-dimensional representation of the color, such as tristimulus values referenced to some illuminant. There are many ways to generate reconstructed spectral distributions from target tristimulus values.[23-41] The solution is not unique; there is an entire "metameric suite" of spectral distributions that share common tristimulus values. The outcome of each reconstruction algorithm differs according to the assumptions and restrictions imposed on the reconstruction.

One way to perform the reconstruction is to identify the unique member of the metameric suite of reflectance curves, $\rho$, with minimum slope squared, integrated over the range of visible wavelengths, i.e.,

$$\min_{\rho} \int_{visible\ \lambda} (d\rho/d\lambda)^2\ d\lambda, \qquad (1)$$

subject to the constraint that $\rho$ must yield the target tristimulus values. Equation 1 was originally proposed by van Trigt.[23-24] Unfortunately, this form of the optimization can sometimes lead to curves that have regions of negative reflectance. To avoid this, the present author has modified the optimization to operate in the space of $z = \ln \rho$, so the new objective function becomes

$$\min_z \int_{visible\ \lambda} (dz/d\lambda)^2\ d\lambda. \tag{2}$$

This keeps $\rho = e^z$ strictly positive. The continuous optimization above can be discretized to make it suitable for numerical solution:

$$\min_z \sum_{i=1}^{n-1} (z_{i+1} - z_i)^2 \tag{3}$$

$$\text{s.t.}\ A_S'\ e^z = XYZ_S.$$

Note that the constraint uses a source illuminant-referenced set of CMFs, $A_S$, which when transposed and multiplied by the $n \times 1$ reflectance vector, $e^z$, equals the target $XYZ_S$. Equation 3 can also be expressed fully in matrix form:

$$\min_z \tfrac{1}{2} z' C z \tag{4}$$

$$\text{s.t.}\ A_S'\ e^z = XYZ_S,$$

where C is an $n \times n$ tridiagonal matrix of finite-differencing constants,

$$C = \begin{bmatrix} 2 & -2 & & & & & \\ -2 & 4 & -2 & & & & \\ & -2 & 4 & -2 & & & \\ & & \ddots & \ddots & \ddots & & \\ & & & & -2 & 4 & -2 \\ & & & & & -2 & 2 \end{bmatrix}. \tag{5}$$

This constrained nonlinear program (NLP) is easily solved by optimization software packages. However, the computational overhead of general-purpose optimization code is unnecessary when solving an NLP with a simple structure such as this one. A far more computationally efficient approach is to apply the method of Lagrange multipliers to this equality constrained NLP. The Lagrangian function is

$$\mathcal{L}(z, \lambda) = \tfrac{1}{2} z'Cz + \lambda'(A_S'\, e^z - XYZ_S), \tag{6}$$

where $\lambda$ is a 3×1 vector of Lagrange multipliers. Setting partial derivatives of the Lagrangian to zero,

$$\begin{aligned}\partial \mathcal{L}/\partial z &= C\,z + \text{diag}(e^z)\, A_S\, \lambda = 0 \\ \partial \mathcal{L}/\partial \lambda &= A_S'\, e^z - XYZ_S = 0,\end{aligned} \tag{7}$$

and solving for $z$ and $\lambda$ yields a stationary point, which for this particular equality-constrained quadratic program is a minimum and a solution. This $(n+3) \times (n+3)$ system of nonlinear equations is readily solved by Newton's method. Defining the vector-valued function, $F$, as

$$F = \begin{Bmatrix} C\,z + \text{diag}(e^z)\, A_S\, \lambda \\ A_S'\, e^z - XYZ_S \end{Bmatrix}, \tag{8}$$

and the Jacobian matrix of first partial derivatives of $F$ as

$$J = \left[\begin{array}{c|c} C + \text{diag}(\text{diag}(e^z)\, A_S\, \lambda) & \text{diag}(e^z)\, A_S \\ \hline A_S'\, \text{diag}(e^z) & 0 \end{array}\right], \tag{9}$$

the change in the values of $z$ and $\lambda$ with each Newton iteration is found by solving the linear system

$$J \begin{Bmatrix} \Delta z \\ \Delta \lambda \end{Bmatrix} = -F. \tag{10}$$

Thus, a reconstructed reflectance curve can be found with a series of linear equation solutions, updating $z$ and $\lambda$ each iteration: $z^{k+1} = z^k + \Delta z$ and $\lambda^{k+1} = \lambda^k + \Delta \lambda$. Convergence is rapid and tests on 140,000 cases show that a solution to within a tolerance of $|F| < 10^{-8}$ is obtained with a mean number of iterations of 6.8.

As an interesting side note, due to the nature of this equality-constrained quadratic program, there is a unique reflectance curve with minimum-slope-squared properties for a given tristimulus value triplet. In other words, there is a one-to-one correspondence between these optimal reflectance curves and their corresponding tristimulus values. Either can be used interchangeably as a canonical representation of a specific color.

Below is a Matlab function that performs the spectral reconstruction. This code also works in the free and open source Octave software.[42] It accepts as input arguments the source-referenced CMFs, `As`, the matrix of finite-differencing constants, `C`, and the source tristimulus values, `XYZ`, and returns the spectral reconstruction, `rho`.

```
function rho=spectral_recon(As, C, XYZ)
% As is nx3 Ws-referenced CMFs, As = diag(Ws)*A
% C is nxn matrix of finite-differencing constants (Eqn 5)
% XYZ is a 3x1 vector of target Ws-referenced tristimulus values
% rho is a nx1 vector of reflectance values (or zeros if failure)
n=size(As,1);rho=zeros(n,1);z=zeros(n,1);lambda=zeros(3,1);count=0;
maxit=20; % max number of iterations
ftol=1.0e-8; % solution tolerance
while count <= maxit
    r=exp(z);
    dr=diag(r);
    v=dr*As*lambda;
    F=[C*z+v; As'*r-XYZ];
    J=[C+diag(v), dr*As; As'*dr, zeros(3)];
    delta=J\(-F); % solve system of equations J*delta = -F
    z=z+delta(1:n);
    lambda=lambda+delta(n+1:n+3);
    if all(abs(F)<ftol)
        rho=exp(z);
        return
    end
    count=count+1;
end
```

Since this code will be executed once for each corresponding color prediction, it is best to precompute $A_S$ and $C$ and pass them as input arguments. The function will fail if improper inputs are provided, such as $XYZ$ outside the spectral locus. The online supplementary documentation provides a worked example of this reconstruction.[43]

## 6. Implementing the Chromatic Adaptation Transform

This section outlines how the spectral reconstruction method can be implemented as a CAT, with Matlab/Octave code snippets included for clarity. It is assumed that the three inputs are the source and destination illuminant white points, $XYZ_S^{wp}$ and $XYZ_D^{wp}$, and the source-illuminant-referenced color, $XYZ_S$.

1. If the illuminant white points provided are scaled so that $Y=100$, divide them by 100 to conform with the tristimulus value scaling convention (0-1) used throughout this presentation. Similarly scale $XYZ_S$ to the 0-1 convention if provided in the 0-100 range.
2. Compute the source and destination illuminant vectors, $W_S$ and $W_D$, from $XYZ_S^{wp}$ and $XYZ_D^{wp}$ using the spectral reconstruction code above, replacing input parameter $A_S$ with the unweighted CMFs $A$:
   ```
   Ws=spectral_recon(A, C, XYZwps);
   Wd=spectral_recon(A, C, XYZwpd);
   ```
   This will produce illuminants that are properly normalized (scalar product $\bar{y}' \cdot W = 1$).
3. Compute illuminant-referenced CMFs $A_S = \overline{W_S} A$ and $A_D = \overline{W_D} A$ (overbar indicates a diagonal matrix created from the specified vector).
   ```
   As=diag(Ws)*A;
   Ad=diag(Wd)*A;
   ```
4. Create the finite-differencing coefficients matrix, $C$, as shown in Equation 5. One way to do this is with the Matlab/Octave statements:
   ```
   C=full(gallery('tridiag',n,-2,4,-2));
   C(1,1)=2;
   C(n,n)=2;
   ```
   where $n$ is the number of rows of matrix $A$.
5. The above steps need to be performed only once for a given source/destination illuminant pair. The remaining steps are done for each corresponding color destination prediction: Compute a reflectance reconstruction from the source color, $XYZ_S$:
   ```
   rho=spectral_recon(As, C, XYZs);
   ```
6. Compute the unadjusted destination color, $XYZ_D = A_D' \rho$:
   ```
   XYZd=Ad'*rho;
   ```
7. Adjust $XYZ_D$ so that it preserves its chromaticity coordinates, $(x, y$ , while having a $Y$ value that matches the $Y$ value of $XYZ_S$. Call this $XYZ_{Dp}$, which is the predicted $W_D$-referenced destination color forming a corresponding color pair with $W_S$.
   ```
   XYZdp=XYZd*XYZs(2)/XYZd(2);
   ```

## 7. Symmetry

In some applications, it is desirable that the CAT be symmetric, that is, if a source color is transformed from illuminant 1 to illuminant 2, and then the destination color obtained is transformed from illuminant 2 back to illuminant 1, the resulting destination color should match the original source color. In the CAT, as proposed above, there may be a small mismatch.

This can easily be corrected, however, by a slight modification to the proposed method. We define a "dual-referenced" CMF as

$$A_{SD} = \overline{W_S}\, \overline{W_D}\, A = \overline{W_D}\, A_S \tag{11}$$

This dual referenced CMF replaces some instances of $A_S$ in the system of nonlinear equations

$$F = \begin{Bmatrix} C\,z + \mathrm{diag}(e^z)\, A_{SD}\, \lambda \\ A_S'\, e^z - XYZ_S \end{Bmatrix} \tag{12}$$

and in the corresponding Jacobian matrix

$$J = \left[ \begin{array}{c|c} C + \mathrm{diag}(\mathrm{diag}(e^z)\, A_{SD}\, \lambda) & \mathrm{diag}(e^z)\, A_{SD} \\ \hline A_S'\, \mathrm{diag}(e^z) & 0 \end{array} \right]. \tag{13}$$

Note that $A_{SD}$ is not used in the second set of equations, as these equations must use $A_S$ to ensure that the reflectance curve has the proper tristimulus values, referenced only to the source illuminant. Justification of this alteration is lengthy and is available in the online supplementary documentation.[44]

Changing the Matlab/Octave code presented earlier for this symmetric case is a simple matter of modifying three lines, as shown in Table 3.

Table 3. Code changes to make the proposed method symmetric.

| Line | Original Code | Symmetric Code |
|---|---|---|
| 1 | `function rho=spectral_recon(As,C,XYZ)` | `function rho=spectral_recon_sym(As,Asd,C,XYZ)` |
| 12 | `v=dr*As*lambda;` | `v=dr*Asd*lambda;` |
| 14 | `J=[C+diag(v),dr*As; As'*dr,zeros(3)];` | `J=[C+diag(v),dr*Asd; As'*dr,zeros(3)];` |

The change to the reflectance curve is typically quite small. For example, using illuminant A as the source and illuminant D65 as the destination, the source color $XYZ_S = (0.2, 0.3, 0.1)$ produces the two reflectance curves shown in Figure 9 using the original spectral reconstruction and the symmetric version.

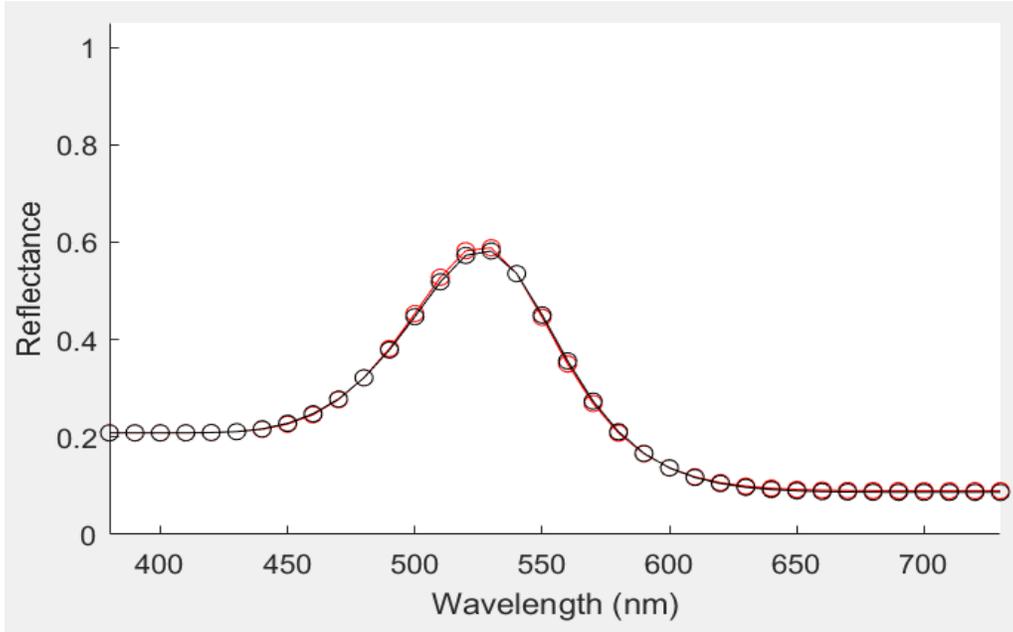

Figure 9. Reconstructed reflectance curve for original proposed method (black) and symmetric version (red).

The two reflectance curves are virtually identical. However, that small change is sufficient to ensure symmetry, as demonstrated in Table 4.

Table 4. Comparison of original and symmetric versions. The fourth column is the outcome of the reversed CAT (D65 to A) applied to the $XYZ_{D65}$ values of the third column.

| Version | $XYZ_A$ | CAT A to D65 $XYZ_{D65}$ | CAT D65 to A $XYZ_A$ |
|---|---|---|---|
| Original | (0.2, 0.3, 0.1) | (0.1707, 0.3000, 0.2426) | (0.2059, 0.3000, 0.1016) |
| Symmetric | (0.2, 0.3, 0.1) | (0.1699, 0.3000, 0.2415) | (0.2000, 0.3000, 0.1000) |

The symmetric version also performs virtually the same as the original proposed method when applied to the corresponding-colors datasets. Table 5 presents color difference values for both methods and the overall means match to two decimal places.

Table 5. Comparison of mean $\Delta E^*_{94}$ color difference for the original proposed method and the symmetric version.

| Dataset | Original Method | Symmetric Version |
| --- | --- | --- |
| CSAJ | 3.72 | 3.70 |
| Helson | 4.10 | 4.09 |
| Lam & Rigg | 3.22 | 3.22 |
| LUTCHI | 4.01 | 4.03 |
| Kuo & Luo | 2.85 | 2.84 |
| Breneman | 5.48 | 5.47 |
| Braun & Fairchild | 4.07 | 4.08 |
| McCann | 9.78 | 9.78 |
| Mean (All Datasets) | 4.74 | 4.74 |
| Mean (No McCann) | 4.01 | 4.01 |

## 8. Robustness

The spectral reconstruction used in the proposed method creates reflectance curves that are strictly positive. As a direct consequence, it is impossible for the proposed CAT to generate a destination color outside the spectral locus, regardless of the source/destination illuminants and source color selected.

To demonstrate this robustness, the proposed CAT (specifically, the symmetric version) is applied to the optimum colors enclosing a $Y=0.3$ slice of the object color solid. Equal energy is used as the source illuminant and a series of destination illuminants with white points located 90% from equal energy to the spectral locus are selected at nine equally spaced angles from equal energy. Figure 10 demonstrates that in all cases, the destination colors remain inside the spectral locus.

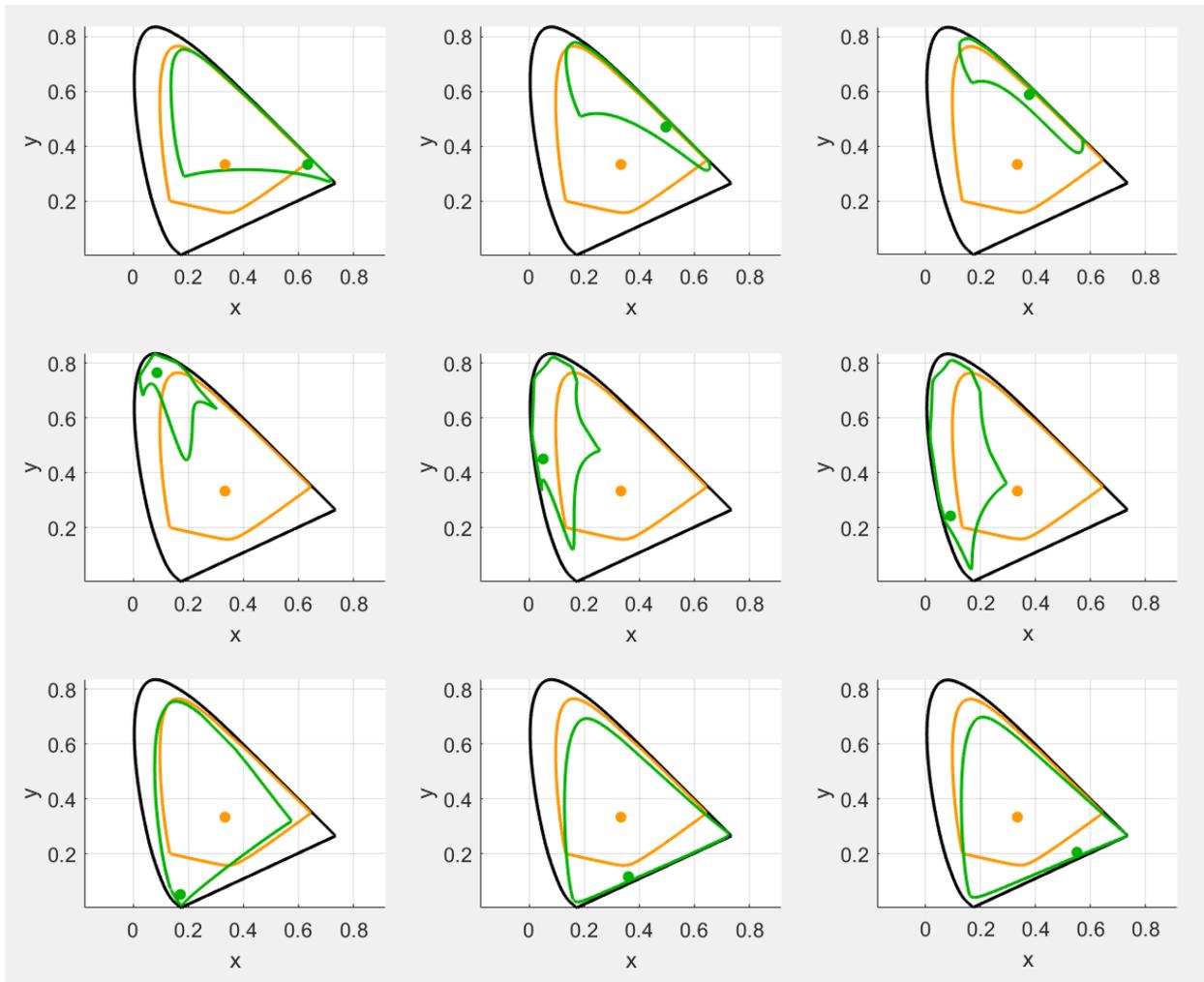

Figure 10. The proposed CAT applied to a $Y=0.3$ slice of the object color solid, using equal energy as the source illuminant (orange) and a variety of destination illuminants that are 90% toward the spectral locus (green).

Compare that to the behavior of CAT16 applied to the same conditions, shown in Figure 11. In all nine cases, there is a portion of the object color solid that is projected outside the spectral locus.

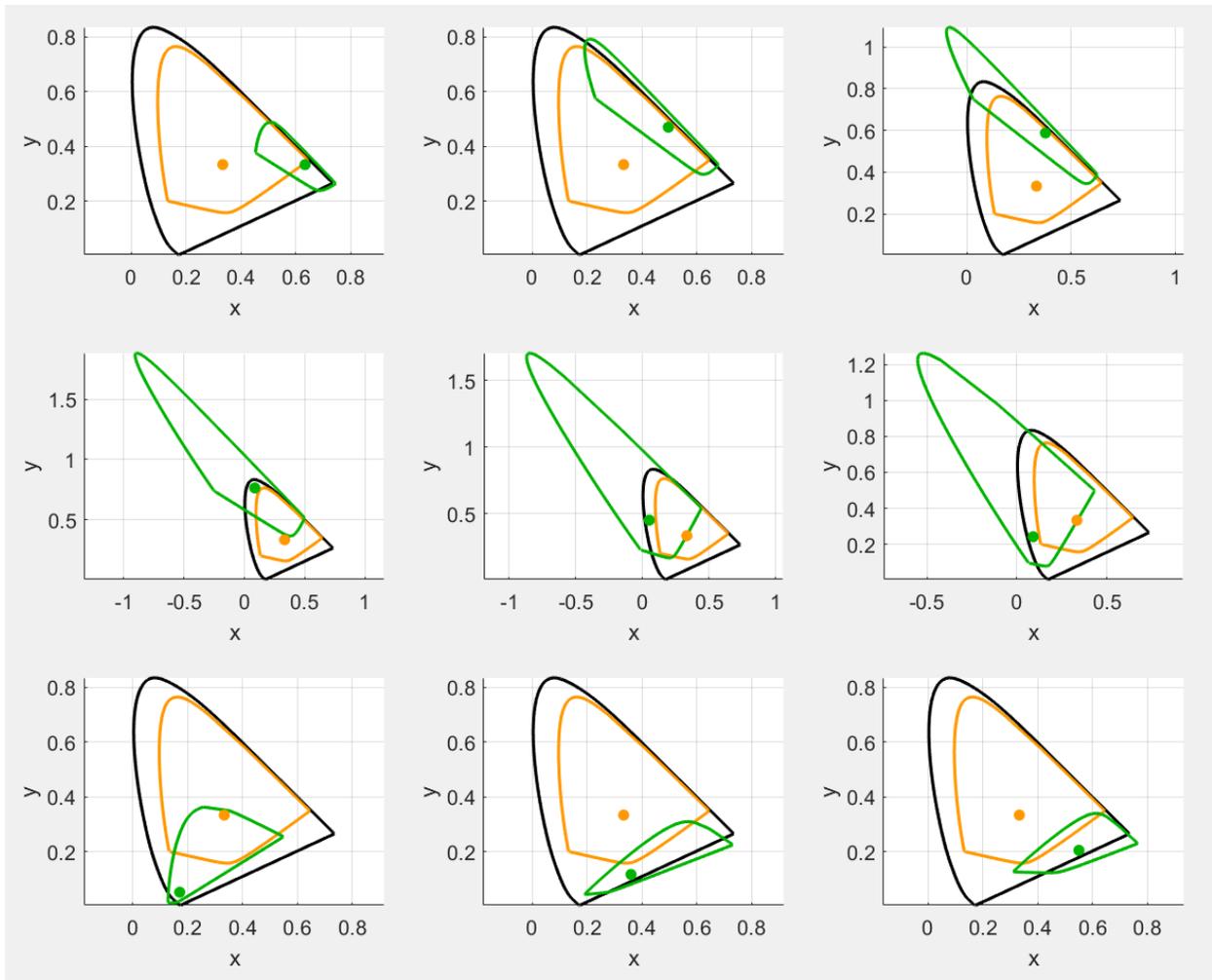

Figure 11. CAT16 applied to the same conditions of Figure 10.

The possibility of a CAT producing negative tristimulus values can be problematic. As Li et al point out in Reference 9, negative tristimulus values can be detrimental to many applications, such as when developing a real surface color gamut as a reference space, or when evaluating image reproduction devices using that reference space. Brill points out in Reference 10, "if CIECAM02 has to be used in imaging applications, any color/illuminant combination should be converted properly and hence some modifications are needed."

The proposed CAT would solve the problem of invalid destination colors. However, the method has two main deficiencies in comparison to existing CATs: increased complexity and increased computational effort. The spectral reconstruction requires solving a series of simultaneous linear equations, whereas existing CATs require only matrix multiplication. Fortunately, implementing the proposed method does not require writing original software for solving the linear systems. The field of numerical linear algebra is mature and public domain software utilities are readily available for solving such systems.[45] These utilities have been available for decades and have been refined over the years to be efficient and robust. They are reliable building blocks for numerical applications such as this.

The other obstacle is computational effort. Linear equation solving requires considerably more computation than simple matrix multiplication, which is all that current CATs require. Based on timed trials of 140,000 cases comparing the execution times of the spectral reconstruction to simple matrix multiplication, the author estimates that a 2-3 order of magnitude increase in computational effort is required (for the case of $n=36$). This is a substantial increase, but if robustness of the CAT is of paramount importance, this increase may be justifiable.

## 9. Degree of Adaptation

Existing CATs implement a parameter, $D$, to address partial adaptation. It is set between 1 and 0 according to the viewing conditions. The diagonal matrix of von Kries coefficients, $\Lambda$, comprising ratios of destination to source illuminant white points, is then parametrically shifted toward the identity matrix as $D$ approaches zero:

$$\Lambda_{adapt} = D\,\Lambda + (1-D)\,I. \tag{14}$$

A similar linear parametric treatment could be applied to the proposed method by adjusting the destination illuminant toward the source illuminant using $D$:

$$W_{D,adapt} = D\,W_D + (1-D)\,W_S. \tag{15}$$

As $D$ approaches zero, the source and destination illuminants become the same and the destination color equals the source color, just as with the $D=0$ case in existing CATs.

The author has not investigated how this partial adaptation treatment compares to the standard treatment. It is a suggested topic for future investigation.

## 10. Conclusions

A proposed CAT has been presented that is based on a spectral reconstruction algorithm, specifically, one that computes the unique spectral distribution within a metameric suite that has minimum slope squared in log space. The reflectance curves thus generated tend to show good similarity to reflectance curves of many object colors found in nature, specifically, commercial paints and natural pigments. The proposed CAT simply computes the effect that changing the illuminant has on tristimulus values computed from the reflectances, using standard CMF calculations. While this approach appears to ignore the phenomenon of adaptation, focusing instead on techniques more closely related to color constancy, the corresponding-color pairs it generates match the empirical corresponding-color datasets as well as or better than existing CATs. The proposed CAT does not use the empirical corresponding-color datasets to "tune" performance like the most recent CATs do.

It was also demonstrated that existing CATs can produce defective results for certain illuminants and source color combinations, whereas the proposed CAT is completely immune to such deficiencies. The price to be paid for this increased robustness is an increase in complexity and computational effort. The additional computation required is quite significant. However, if robustness of the CAT is of primary importance, such as with certain imaging applications, then this increase may be justifiable.